\documentclass{aa}
\usepackage{graphicx}
\usepackage{psfig}
\newcommand{\mincir}{\raise
  -2.truept\hbox{\rlap{\hbox{$\sim$}}\raise5.truept \hbox{$<$}\ }}
\newcommand{\magcir}{\raise
  -2.truept\hbox{\rlap{\hbox{$\sim$}}\raise5.truept \hbox{$>$}\ }}
\newcommand{\hm}{\,h^{-1}{\rm Mpc}}
\begin{document}


\title{Simulating the Soft X-ray excess in clusters of galaxies} 

\author{L.-M. Cheng\inst{1,2,3,4} \and S. Borgani\inst{1,4}
        \and P. Tozzi\inst{5} \and L. Tornatore\inst{1,4} \and
        A. Diaferio\inst{6} \and K. Dolag\inst{7} \and
        X.-T. He\inst{2} \and
        L. Moscardini\inst{8} \and G. Murante\inst{9} 
\and G. Tormen\inst{10}}

\offprints{L.-M. Cheng}
\mail{cheng@ts.astro.it}
\institute{Dipartimento di Astronomia dell'Universit\'a di Trieste,
via Tiepolo 11, I-34131 Trieste, Italy  
\and 
  Department of Astronomy, Beijing Normal University,  
  Beijing 100875, China
\and 
 National Astronomical Observatories, Chinese Academy of Sciences, 
 Beijing 100012, China
\and 
  INFN-National Institute for Nuclear Physics, Trieste, Italy
\and 
INAF, Osservatorio Astronomico di Trieste, via Tiepolo 11,
 I-34131 Trieste, Italy 
\and
Dipartimento di Fisica Generale ``Amedeo Avogadro'', Universit\'a
  degli Studi di Torino, Torino, Italy 
\and
Max-Planck-Institut f\"ur Astrophysik, Karl-Schwarzschild Strasse
  1, Garching bei M\"unchen, Germany 
\and
Dipartimento di Astronomia, Universit\`a di Bologna, via Ranzani
  1, I-40127 Bologna, Italy 
\and
INAF, Osservatorio Astronomico di Torino, Strada Osservatorio 20,
  I-10025 Pino Torinese, Italy 
\and
Dipartimento di Astronomia, Universit\`a di Padova, vicolo
  dell'Osservatorio 2, I-35122 Padova, Italy 
} 


   \titlerunning{Soft X-ray excess}
   \authorrunning{Cheng et al.}

\abstract{The detection of excess of soft X-ray or Extreme Ultraviolet
(EUV) radiation, above the thermal contribution from the hot
intracluster medium (ICM), has been a controversial subject ever since
the initial discovery of this phenomenon. We use a large--scale
hydrodynamical simulation of a concordance $\Lambda$CDM model, to
investigate the possible thermal origin for such an excess in a set of
20 simulated clusters having temperatures in the range 1--7 keV.
Simulated clusters are analysed by mimicking the observational
procedure applied to ROSAT--PSPC data, which for the first time showed
evidences for the soft X--ray excess: we compare the low--energy
(e.g. [0.2--1] keV) part of the spectrum of each cluster with that
predicted for a plasma having temperature and metallicity computed
after weighting by emissivity in a harder band (e.g., [1--2] keV). For
cluster--centric distances $0.4< R/R_{\rm vir}< 0.7$ we detect a
significant excess in most of the simulated clusters, whose relative
amount changes from cluster to cluster and, for the same
cluster, by changing the projection direction. In about 30 per cent of
the cases, the soft X--ray flux is measured to be at least 50 per cent
larger than predicted by the one--temperature plasma model. We find
that this excess is generated in most cases within the cluster
virialized regions. It is mainly contributed by low--entropy and
high--density gas associated with merging sub--halos, rather than to
diffuse warm gas. Only in a few cases the excess arises from
fore/background groups observed in projection, while no evidence is
found for a significant contribution from gas lying within
large--scale filaments. We compute the distribution of the relative
soft excess, as a function of the cluster--centric distance, and
compare it with the observational result by Bonamente et al. (2003)
for the Coma cluster. Similar to observations, we find that the
relative excess increases with the distance from the cluster center,
with no significant excess detected for $R<0.4R_{\rm vir}$. However,
an excess as large as that reported for the Coma cluster at scales
$0.4\mincir R/R_{\rm vir}\mincir 0.7$ is found to be rather unusual in
our set of simulated clusters.

}

\maketitle
\section{Introduction}
Clusters of galaxies provide a reservoir of baryons in the form of a
hot plasma with typical temperatures of $10^7$--$10^8$ K, which emits
over a broad band from Extreme Ultraviolet (EUV) to $\sim 10$ keV
X--rays. Most of the observed X-ray features of clusters can be well
accounted for within the framework of thermal bremsstrahlung emission
plus emission lines associated to the metal content of the
intra--cluster medium (ICM).  However, a number of observations with
the Extreme Ultraviolet Explorer (EUVE; e.g. Lieu et al. 1996a,b;
Mittaz et al. 1998; Maloney \& Bland-Hawthorn 2001), ROSAT
(e.g. Bonamente et al. 2001a,b) and XMM-Newton (e.g. Finoguenov et
al. 2003; Kaastra et al. 2003) have claimed the detection of an excess
of EUV and soft X-ray emission in the spectrum of several clusters,
with respect to what expected from a one--temperature plasma model.

A number of suggestions have been proposed for the origin of this
excess, the two most popular scenarios being the non--thermal origin
from inverse Compton scattering of the cosmic microwave background
photons by relativistic electrons in the intracluster gas (e.g. Hwang
1997; En$\beta$lin \& Biermann 1998; De Paolis et al. 2003), and the
thermal origin due to warm gas at $T\sim 10^6\,K$, either from inside
clusters or from diffuse filaments outside clusters (Lieu et
al. 1996b; Nevalainen et al. 2003).  However, the existence of the
excess has been disputed by Bowyer et al. (1999), who argued that the
EUV excess is an artifact caused by improper subtraction of the
instrumental background (see also Berghofer \& Bowyer 2002; Durret et
al. 2002), although they concluded that a relatively weak EUV excess
in the Virgo and Coma clusters may be real. Bregman et al. (2004) also
argued that the excess may be caused by an improper inclusion in the
data analysis of the effect of fluctuations of the galactic hydrogen
column density.

Even within the framework of the thermal models, it is a matter of
debate whether the gas responsible for the excess is located within
clusters or in large--scale filamentary structures. For instance,
Ettori (2003) found evidence for as much as $17$ per cent of the
baryons in clusters to be presumably in the form of warm
($10^5$--$10^7$ K) material, which must emit EUV or soft X-ray photons
in excess to those expected from the hot phase of the ICM. On the
other hand, Kaastra et al. (2003) and Finoguenov et al. (2003) claim
that the soft excess may have originated in filaments in the gas
distribution in the vicinity of clusters.

In this paper, we present an analysis of a set of clusters extracted
from a large cosmological hydrodynamical simulation (Borgani et
al. 2004, Paper I), which is aimed at investigating the presence and
origin of a soft X--ray excess in their spectra. By taking advantage
of the cosmological environment of our simulation, which includes
radiative cooling, star formation and galactic winds triggered by
supernova (SN), we ``observe'' clusters in projection and estimate
their spectra also including the contribution from the
background/foreground large--scale gas distribution. Since in our
simulation we treat only thermal emissivity processes, the two main
questions that we intend to address are the following: (a) does a
realistic description of the evolution of cosmic baryons account for a
soft X--ray excess of thermal origin as large as that observed in the
spectra of clusters?  (b) is the excess associated to warm gas
residing within clusters or to large--scale filaments observed in
projection?

The structure of the paper is as follows. In Section 2 we describe our
set of simulated clusters. After describing the procedure to compute
the synthetic spectra, we present in Section 3 our results on the
excess, its origin and a comparison with observations. We draw our
main conclusions in Section 4.

\section{The simulated clusters}
We analyze a representative set of 20 simulated clusters, which are
extracted from a cosmological box for a standard flat $\Lambda$CDM
cosmological model, with $\Omega_m=0.3, \Omega_\Lambda=0.7$, Hubble
constant $H_0=100h$ km s$^{-1}$Mpc$^{-1}$,$h=0.7$, baryon density
$\Omega_b=0.04$ and $\sigma_8=0.8$ for the normalization of the power
spectrum. The box has side-length of $192h^{-1}\rm Mpc$ and contains
$480^3$ dark matter particles and an initially equal number of gas
particles, thus resulting in $m_{\rm DM}=4.6\times 10^9 h^{-1}M_\odot$ and
$m_{\rm gas}=6.9\times 10^8 h^{-1}M_\odot$ for the mass of the two particle
species. The Plummer--equivalent gravitational force softening is set
to $7.5\,h^{-1}$kpc in physical units from $z=2$ to $z=0$, while it is
kept fixed in comoving units at higher redshift. The SPH smoothing
scale is allowed to decrease at most to one--fourth of the
gravitational softening (see Paper I, for a more
detailed description of this simulation). 

The run has been evolved using {\tt P-GADGET2}, a massively parallel
Tree--SPH code (Springel et al. 2001) with fully adaptive time-step
integration. The implementation of SPH adopted in the code follows the
entropy--conserving formulation by Springel \& Hernquist (2002). The
simulation includes a treatment of star formation based on a
sub--resolution model of the interstellar medium and the effect of
galactic winds powered by SN-II explosions (Springel \& Hernquist
2003, SH03). The code also includes a treatment of metal production
from SN-II. The resulting metallicity value, which is assigned to each
gas and star particle, has to be interpreted as a global value which
is contributed by different heavy elements with solar relative
abundances.  We exclude those particles from the computation of X--ray
emissivity having temperature below $3\times 10^4$ K and gas density
above $500\bar\rho_{\rm bar}$, being $\bar\rho_{\rm bar}$ the mean
baryon density. Furthermore, following SH03, each gas particle which
lies above a limiting density threshold is assumed to be composed of a
hot ionized phase and a cold neutral phase, whose relative amounts
depend on the local conditions of density and temperature. Since such
particles are aimed at describing the multi-phase nature of the
inter-stellar medium, we decided to exclude also their contribution in
the computation of the ICM X--ray emissivity. Although the number of
such particles is always very small, their high density may cause a
sizable, although spurious, contribution to the soft X--ray emission.

\begin{table*}
\caption{\footnotesize The set of simulated clusters. Column 1:
Cluster identification; Column 2: virial mass; Column 3: virial
radius; Columns 4 and 5: temperature and metallicity within $R_{\rm
vir}$, emission--weighted in the [1--2] keV energy band; Columns
6--8: relative soft excess (see eq. \ref{eq:relex}) in the [0.2--1]
keV band, computed in the region $0.4<R/R_{\rm vir}<0.7$, by
projecting along three orthogonal directions.}
\begin{center}
\begin{tabular}{cccccccc}
\hline
Cluster& $M_{\rm vir}$ & $R_{\rm vir}$& $T_{\rm ew}$ & $Z_{\rm ew}$ & $\eta_x$ & $\eta_y$ & $\eta_z$ \\
Index & $[10^{14}h^{-1}M_{\odot}]$ & $[h^{-1}Mpc]$& $[keV]$ & $[Z_{\odot}]$&  &  &  \\
\hline
CL01 &  1.60& 1.11 & 2.43 & 0.16 &0.97  & 0.19 & 0.10\\
CL02 &  2.46& 1.28 & 2.74 & 0.17 &0.12  & 0.80 & 0.53\\
CL03 &  2.59& 1.30 & 3.05 & 0.17 &0.17  & 0.18 &0.21\\
CL04 &  7.00& 1.82 & 5.12 & 0.13 &0.11  & 0.40 &0.10\\
CL05 &  2.00& 1.20 & 2.42 & 0.19 &0.48  & 0.20 &0.21\\
CL06 &  1.72& 1.14 & 2.40 & 0.18 &0.17  & 0.17 &0.19\\
CL07 &  1.99& 1.20 & 2.59 & 0.14 &0.28  & 0.23 &0.24\\
CL08 &  13.0& 2.23 & 6.50 & 0.11 &0.58  & 0.26 &0.36\\
CL09 &  2.57& 1.30 & 3.02 & 0.12 &0.53  & 0.44 &0.11\\
CL10 &  3.76& 1.48 & 3.74 & 0.11 &0.34  & 0.31 &0.17\\
CL11  &1.46  & 1.08 & 2.02 &0.18&0.56  &  0.30 &0.68\\
CL12  &1.57  & 1.21 & 1.56 &0.24&0.67  &  0.61 &0.26\\ 
CL13  &1.08  & 0.97 & 1.87 &0.20&0.22  &  1.47 &1.34\\ 
CL14  &1.45  & 1.07 & 2.17 &0.13&1.73  &  1.36 &1.68\\ 
CL15  &2.07  & 1.21 & 2.60 &0.16&0.17  &  0.18 &0.12\\ 
CL16  &1.76  & 1.15 & 1.79 &0.26&0.60  &  0.70 &0.26\\
CL17  &6.04  & 1.81 & 5.14 &0.12&0.09  &  0.07 &0.34\\ 
CL18  &1.06  & 1.04 & 1.21 &0.26&1.79  &  1.19 &1.22\\
CL19  &3.34  & 1.41 & 3.50 &0.15&0.17  &  0.20 &0.13\\
CL20  &2.90  & 1.42 & 2.58 &0.11&0.57  &  0.53 &0.50\\ 

\hline\\[-0.6em]
\end{tabular}
\end{center}

\end{table*}


The selected clusters have virial masses spanning about a decade from
$\sim 10^{14}h^{-1}M_\odot$ to $\sim 10^{15}h^{-1}M_\odot$ (see Table
1). We did not apply any particular criterion to select the clusters
to be analysed and, therefore, our set is representative of the whole
cluster population in our simulation within this mass range. For each
cluster, we measure the virial radius, $R_{\rm vir}$, as the distance
from the most--bound DM particle, which encompasses an average density
equal to the virial density for our cosmological model (e.g., Eke et
al. 1996). Accordingly, the virial mass, $M_{\rm vir}$, is defined as
the mass contained within $R_{\rm vir}$. We refer to Paper I for a
description of the cluster identification algorithm that we have
applied. The emission--weighted temperature and metallicity in a given
energy band used to model the hot ICM also are listed in Table 1.
Around each cluster we extract a spherical region extending out to
$6\,R_{\rm vir}$. This region is then observed in projection, by
extracting a cylinder with axis on the center of each cluster. This
allows us to account for the contribution of the surrounding
large--scale structure to the X--ray emission of each cluster. As we
will show below, taking the fore/back--ground structure out to
$6R_{\rm vir}$ is sufficient to obtain converged estimates of this
contribution.

\section{Analysis and Results}

\subsection{Measuring the Soft Excess}

The X--ray luminosity contributed by the $i$--th gas particle in the
simulation is computed according to
\begin{equation}
L_{X,i}=(\mu m_p)^{-2}
\left(\frac{n_e}{n_H}\right)_i
m_i\rho_i\Lambda(T_i,Z_i;E_1,E_2) 
\end{equation}
where $m_i$ and $\rho_i$ are the mass and the density of the hot phase
of that particle, respectively, $\mu$ is the mean molecular weight,
$n_e$ and $n_H$ are the number densities of electrons and protons,
respectively. The cooling function
$\Lambda(T,Z;E_1,E_2)$ is calculated within the energy band
[$E_1$--$E_2$] using the plasma emission model by Raymond \& Smith
(1977), where $Z$ is the gas metallicity.  Using this cooling
function, the spectra are computed by binning the emissivity within
energy intervals, so as to have an energy resolution $\Delta \log
E=0.01$.

In their analysis of the soft X--ray excess of the Coma cluster from
ROSAT--PSPC data, Bonamente et al. (2003) apply a MEKAL model to fit
the high--energy ([1--2] keV) portion of the spectrum. Then they
extrapolate the best--fitting model to a lower energy ([0.2--1] keV) band,
where the predicted spectrum is compared with the actually observed
one. In order to reproduce this same procedure, we would be
required to simulate mock observations of our simulated clusters, and
extract a spectrum with signal--to--noise appropriate for a realistic
exposure time and using the appropriate response function. Since
reproducing in detail the observational setup is beyond the scope of
this paper, we adopt the approach of computing for each cluster the
emission--weighted temperature,
\begin{equation}
T_{\rm ew}=\frac{\sum_i m_i \rho_i \Lambda(T_i,Z_i;E_1,
E_2)T_i}{\sum_i m_i \rho_i \Lambda(T_i,Z_i;E_1, E_2)}
\label{eq:tew}
\end{equation}
and the emission--weighted metallicity,
\begin{equation}
Z_{\rm ew}=\frac{\sum_i m_i \rho_i \Lambda(T_i,Z_i;E_1,
E_2)Z_i}{\sum_i m_i \rho_i \Lambda(T_i,Z_i;E_1,E_2)}.
\label{eq:zew}
\end{equation}
in the [1--2] keV band. Then, we compute the corresponding MEKAL spectrum
for this temperature and metallicity, to be compared with the actual
synthetic spectrum in the [0.2--1] keV band (we use in the following
the solar photospheric abundance by Anders \& Grevesse 1989, when
expressing metallicity in solar units). Clearly, our approach is
correct as long as emission--weighted measures coincide with the
corresponding quantities obtained from spectral fitting. Mazzotta et
al. (2004) have recently pointed out that the complex ICM
temperature structure in simulated clusters causes the
spectral--fitting temperatures to be about 20 per cent lower than
$T_{\rm ew}$ (see also Mathiesen \& Evrard 2001). They provide an
expression for a spectroscopic--like temperature, which represents an
effective recipe to compute the actual spectral temperature in
simulations. However, the fitting function given by Mazzotta et al.
is specific to the response function and energy coverage of the
Chandra and XMM--Newton detectors, therefore not applicable to the
ROSAT--PSPC. Furthermore, no such effective recipe has been
calibrated so far to estimate the spectroscopic metallicity in
simulated clusters. Finally, the spectroscopic--like temperature is
well defined only for clusters hotter than 3 keV, while a significant
fraction of the simulated clusters in our set has lower temperature
(see Table 1). For these reasons, we prefer to use here
eqs.(\ref{eq:tew}) and (\ref{eq:zew}) for the definition of
temperature and metallicity. We defer to a forthcoming paper a more
detailed analysis, based on synthetic XMM-Newton pn and MOS spectra of
simulated clusters, to investigate the detectability of a soft excess
under realistic observing conditions (Cheng et al., in preparation).

The values of $T_{\rm ew}$ and $Z_{\rm ew}$ in the [1--2] keV band are
given in Table 1 for all our clusters. The resulting metallicity
values are smaller than the typical observed Fe abundance, $\sim 0.3
Z_{\odot}$ (e.g., Arnaud et al. 2001; Ikebe et al. 2002; Baumgartner
et al. 2003). A careful study of the ICM metallicity would require
including in simulations the contribution of the SN-Ia and accounting
for the different yields of different elements (e.g., Tornatore et
al. 2004). In principle, this may represent a limitation of the
present analysis, since metal lines are expected to give a significant
contribution to the total emissivity in the soft part of the spectrum,
where we are seeking for the excess.  In order to test whether our
final results are affected by the uncertain description of the ICM
metal enrichment, we have verified by how much our results change when
we assume either $Z=0$ or $Z=0.25 Z_{\odot}$ in the MEKAL spectrum of
a plasma with temperature $T=2$ keV. We find that the contribution to
the total emission from metal lines turns out to be $\sim 3$ per cent
in the [0.2--1] keV band, this tiny difference being due to the lack of
prominent lines in the above energy range. Therefore, we expect that
our final results should be largely insensitive to the approximate
treatment of the ICM metal enrichment.

Having estimated $T_{\rm ew}$ and $Z_{\rm ew}$ for each cluster by
emission-weighting in the [1--2] keV band, we can now compute the
corresponding one--temperature and one--metallicity spectrum. This
spectrum is then compared with the synthetic one, obtained by summing the
contributions from all gas particles within the ``observational''
cylinder of each cluster. The presence and amount of a soft excess is
then established by comparing the two spectra in the [0.2--1] keV band.

\begin{center}
\begin{figure*}
\psfig{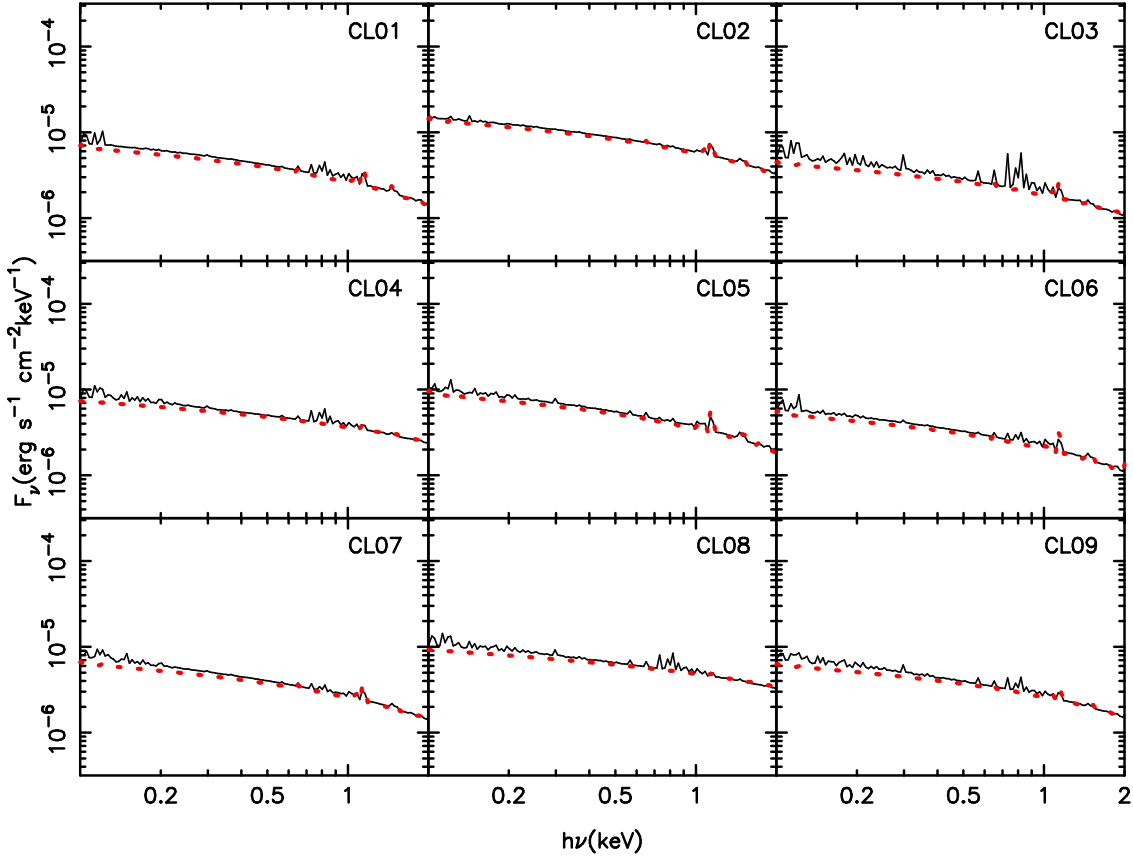}
\caption{Simulated spectra as seen in the projected region within $0.4
R_{\rm vir}$ from the cluster center, for a representative subset of
9 clusters (solid lines). The dotted lines are for the spectra expected
from a one-temperature and one-metallicity plasma model, using the
values of $T_{\rm ew}$ and $Z_{\rm ew}$ computed within the same region in the
[1--2] keV energy band.}
\label{fi:sp04}
\end{figure*}
\end{center}

The result of this comparison is shown in Figure \ref{fi:sp04}, where
we plot the projected luminosity density as a function of the
frequency, $F_\nu$, for cluster-centric distances $R\le 0.4\,R_{\rm
vir}$. Instead of showing spectra for the whole cluster set, we report
here the results only for the first nine clusters of the list. The
solid lines show the synthetic spectra, while the dotted lines are the
prediction from the one--temperature and one--metallicity model. Quite
apparently, the two spectra are always very similar, thus
demonstrating that no appreciable soft excess is generated in the
central regions of our simulated clusters. The only appreciable
difference is related to the presence of soft metal lines in the
synthetic spectra.
%

\begin{center}
\begin{figure*}
\psfig{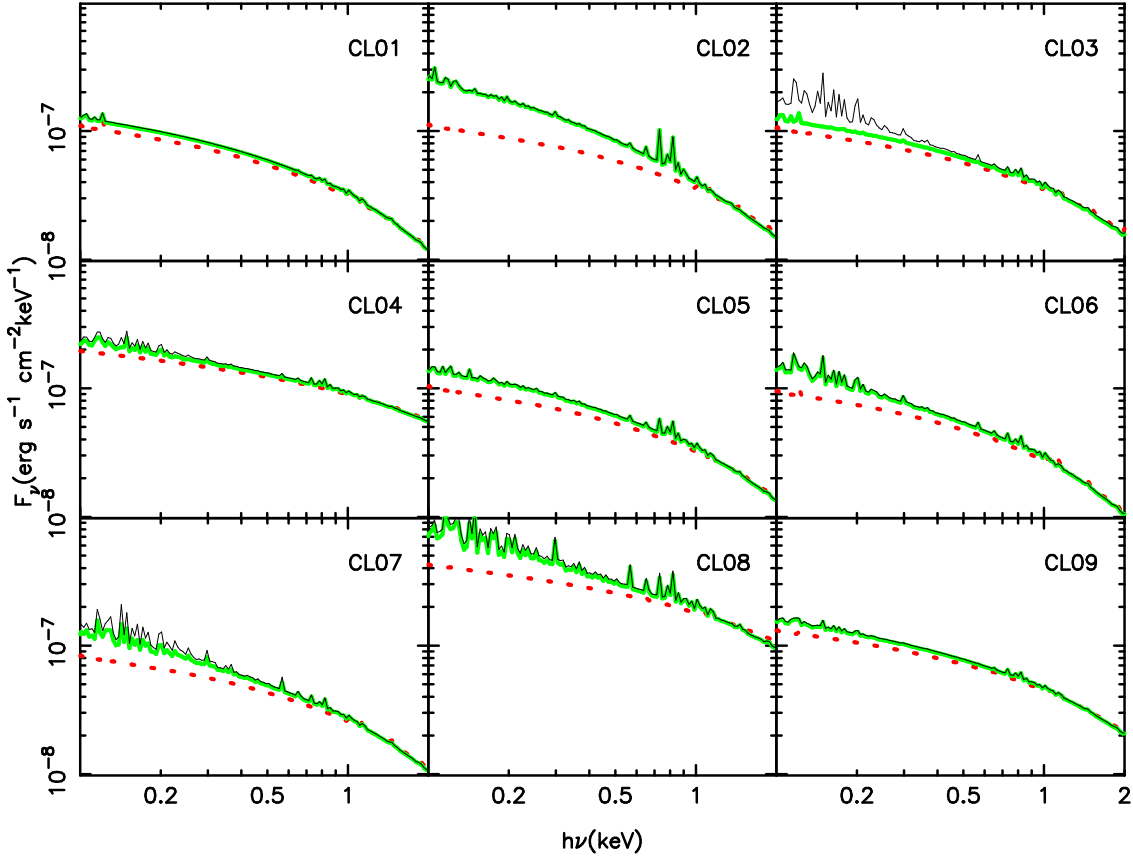}
\caption{The same as in Figure \ref{fi:sp04}, but in the 0.4--0.7
  $R_{\rm vir}$ projected region. In addition the thick lines are the
  same as the thin lines, but using only the gas particles whose
  line-of-sight distance from the cluster centers is $\le R_{\rm
    vir}$.}
\label{fi:sp07}
\end{figure*}
\end{center}

A quite different result is found if we concentrate, instead, on more
external cluster regions, $0.4\le R/R_{\rm vir}\le 0.7$ (see Figure
\ref{fi:sp07}), which corresponds to the typical scales where the soft
excess is detected from observational data. In this case, a fair
number of clusters in our set shows evidence for a significant excess
at $h\nu\mincir 1$ keV, thus witnessing the presence of a warm gas
component, which contributes to the soft X--ray emission. This result
is consistent with the observational evidence for a
X--ray excess, that is more pronounced at larger cluster--centric
radii (e.g., Bonamente et al. 2002; see the discussion here below). 

In order to quantify the amount of soft excess, we define the relative
excess 
\begin{equation}
\eta=\frac{p-q}{q}
\label{eq:relex}
\end{equation}
(see Bonamente et al. 2002), where $p$ is the flux in a given soft
band, as computed from the synthetic spectrum, and $q$ is the
prediction from the hot ICM model. We provide in Table 1 the values of
the relative excess measured in this region, after projecting each
cluster along three orthogonal directions. In about 30 per cent of the
cases, we find a relative excess $\eta >0.5$. In several cases, the
$\eta$ value for the same cluster changes quite substantially with the
projection direction. This already indicates that the excess is
sensitive to the occurrence of a few significant structures along the
line-of-sight, rather than to the overall large--scale distribution of
the gas.

Having established that the soft excess phenomenon is rather common in
the outskirt region of simulated clusters, when observed in
projection, it is worth asking whether such an excess is generated by
warm gas inside clusters or is associated to fore/back--ground
large--scale filamentary structures, extending over scales of several
Mpc.  Large-scale hydrodynamic simulations (e.g. Cen \& Ostriker 1999;
Cen et al. 2001; Dav\'e et al. 2001) indicate that about half of the
baryons in the local Universe exist in the form of a warm-hot
intergalactic medium (WHIM) with temperature $ T\sim 10^5$--$10^7
K$. The WHIM permeates the so-far elusive large--scale filamentary
structures, from which galaxy clusters continuously accrete gas.
Therefore, it is quite possible that the soft excess originates from
large--scale diffuse filaments. On the other hand, an observational
census of the baryons inside clusters (Ettori 2003) suggests that
about 20 per cent of the ICM may be contributed by warm gas in the
above temperature range. This gas would also provide a soft X--ray
emissivity on top of that generated by the hot gas heated to the
cluster virial temperature.

Tracing in simulations the gas distribution in the large--scale
environment of galaxy clusters offers the possibility to distinguish
between the internal and the external origin for the soft excess. To
this purpose, we compute the synthetic spectrum of each cluster by
excluding from the observational cylinder all the gas particles lying
at a distance larger than $R_{\rm vir}$ along the line-of-sight from
the cluster center. The corresponding spectra are shown with the thick
lines in Figure \ref{fi:sp07}. In the large majority of the cases we
find that, after excluding the contribution from the gas outside
$R_{\rm vir}$, the spectrum shows only a very modest change. This
implies that the soft excess is mostly contributed by warm gas
residing inside the virial cluster regions, while excluding a
significant contribution from the large--scale filaments. The only
exception, among the nine clusters shown in Fig. \ref{fi:sp07}, is
represented by CL03, for which a significant part of the excess is
generated outside the virial region. A visual inspection of the gas
distribution along the projection direction reveals the presence of a
small group just approaching the virial region of the cluster, which
is responsible for this excess.

\begin{figure}
\psfig{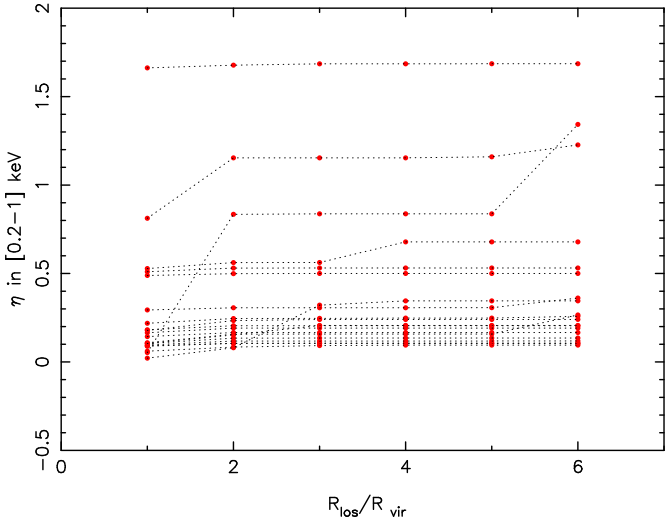}
\caption{The relative excess $\eta$ in the [0.2-1] keV energy band,
  estimated in the region $0.4<R/R_{\rm vir}<0.7$, as a function of
  the maximum line-of-sight distance from the cluster center, $R_{\rm
  los}$. Here we show results for the 20 clusters projected only along
  one direction.}
\label{fi:exdis}
\end{figure}

In order to investigate this point further, we show in Figure
\ref{fi:exdis} how the relative excess changes with the extension
along the line-of-sight, $R_{\rm los}$, of the region where the
spectra are computed. In most cases the values of $\eta$ do not
increase beyond $R_{\rm los}=R_{\rm vir}$, thus confirming that the
excess does not generally receive a significant contribution outside
the cluster virial regions.  In the few cases where the excess
increases along the line-of-sight, this takes place always in a narrow
interval of distance. This confirms that the excess is contributed by
individual small--scale structures (i.e. fore/background groups),
rather than by the integrated effect of large--scale filaments.

In general, the amount of soft X--ray emission from filaments may depend
on the energy feedback from SN and AGN, that heats the diffuse
baryons: an efficient feedback brings the gas on a high adiabat, thus
preventing it from reaching the density contrast of dark matter within
filaments, therefore suppressing its X--ray emissivity. As discussed
in Paper I, the feedback used in this simulation may be somewhat too
weak to prevent overcooling within clusters and to break to the
observed level the self--similarity of X-ray scaling relations at the
scale of galaxy groups. If more efficient feedback needs to be
introduced, then we expect an even smaller contribution of filaments
to the soft X--ray excess.

\begin{center}
\begin{figure*}
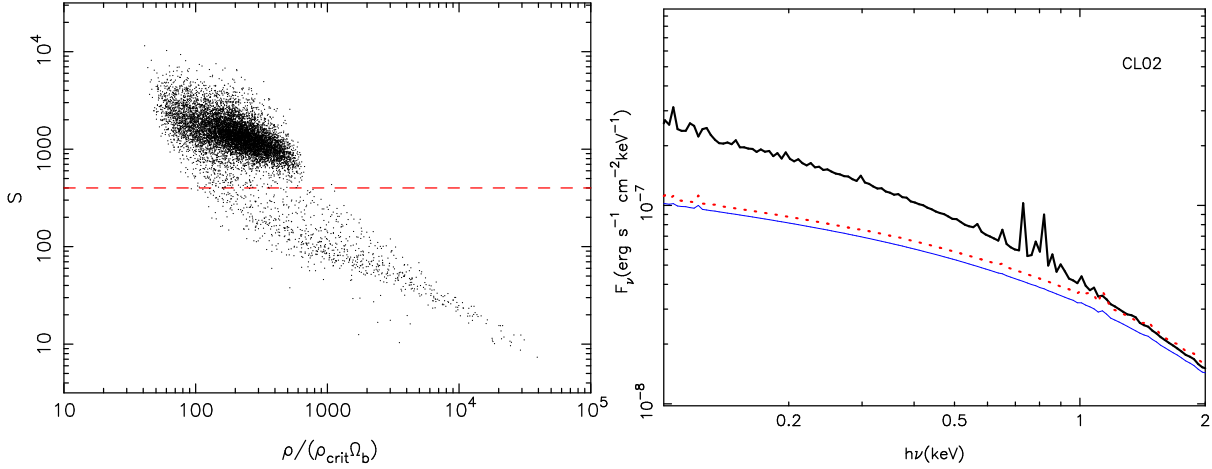

\hbox{\psfig{file=phase.ps,width=8cm,angle=270}\psfig{file=25174.gt.s400.ps,width=8cm,angle=270}} 
\caption{Tracing the gas particles responsible for the soft excess in
  the CL02 cluster. Left panel: the phase diagram shows entropy
  vs. density for the gas particles at projected cluster--centric
  distance $0.4<R/R_{\rm vir}<0.7$. Right panel: the synthetic
  spectrum of CL02 (thick line) is compared with that expected from
  the hot ICM one (dotted curve) and with the synthetic one obtained by
  excluding from the computations all the gas particles with entropy
  $S<400$ keV cm$^2$ (horizontal dashed line in the left panel).}
\label{fi:phase}
\end{figure*}
\end{center}

In order to verify whether the soft excess is generated by a diffuse
warm gas phase or by high density gas within subhalos, we plot in the
left panel of Figure \ref{fi:phase} the entropy--density phase diagram
of the gas lying at projected cluster--centric distance $0.4<R/R_{\rm
vir}<0.7$ for the CL02 cluster. This structure is the one displaying
the largest excess among those shown in Fig.\ref{fi:sp07}.  The
entropy of the $i$--th gas particle is defined as
$S_i=T_i/n_{e,i}^{2/3}$, where $T_i$ is its temperature (in keV) and
$n_{e,i}$ is the associated number density of electrons (in
cm$^{-3}$). Quite apparently, gas particles occupy two distinct
regions of the $S$--$\rho$ plane; (1) the high--entropy low--density
region, which corresponds to the shocked phase formed by gas whose
high entropy has been generated by diffuse accretion; (2) a tail of
condensed gas at lower entropy and high density, which is formed by
gas within merging subhalos that preserved its low entropy during the
accretion phase. As shown in the right panel of Fig. \ref{fi:phase},
the soft excess disappears once we remove the condensed gas phase
(i.e., particles with $S<400$ keV cm$^2$) from the computation of the
spectrum. This demonstrates that the soft excess is associated with the
presence of previously virialized clumps of high density gas at a
temperature of a few tenths of keV, which are still surviving in the
ICM, rather than to a diffuse phase of warm gas superimposed on the
hot cluster atmosphere.

\subsection{Comparison with observations}

\begin{center}
\begin{figure}
\psfig{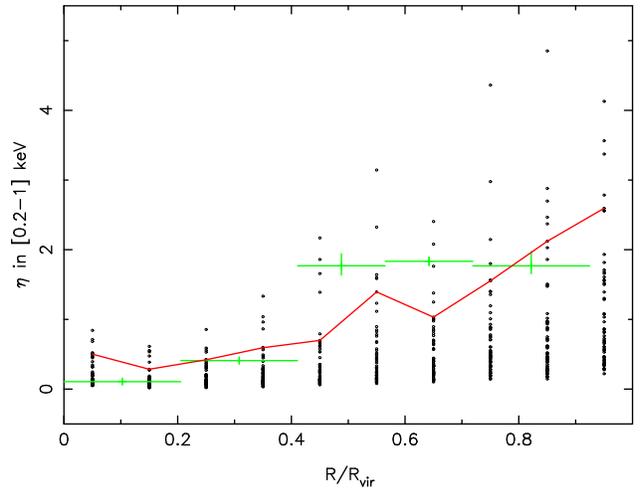}
\caption{The relative excess $\eta$ in the [0.2-1] keV energy band, as
  a function of the projected cluster--centric distance, in units of
  the virial radius, from the simulated clusters (points), compared to
  the observational results for Coma (crosses with errorbars) as
  reported by Bonamente et al. (2003).  With the exception of the
  outermost point, each observational point represents the average
  excess value computed over four quadrants, in which the Coma cluster
  has been divided. For each simulated cluster, we report the excess
  computed by projecting it along three orthogonal directions. The solid
  line shows the upper $90$ percentile for the distribution of
  the excess computed over the ensemble of simulated clusters.}
\label{obs_coma}
\end{figure}
\end{center}

Using ROSAT--PSPC pointings of the Coma cluster, Bonamente et
al. (2003) reported the detection of a soft excess in the [0.2--1] keV
band at a high statistical significance. After performing the analysis
at different angular distances from the cluster center,
they concluded that the relative excess is an increasing function of
this distance. In order to verify how typical is this result within
our set of simulated clusters, we convert angular scales at
the redshift of Coma to physical scales, and then rescale physical
scales in units of the virial radius. To this purpose, we take the
value, $R_{\rm vir}=1.9\hm$ obtained for the Coma cluster by Lokas \&
Mamon (2003) from their analysis of the kinematics of cluster galaxies.
The result of this comparison is shown in Figure
\ref{obs_coma}. Crosses are the observational values of the average
excess within different annuli, as reported in Table 2 of Bonamente
et al. (2003).  For each of the 20 simulated clusters, we plot the
excess obtained by projecting along three orthogonal directions. The
solid line marks the upper 90 percentile in the distribution of the
relative soft excesses in simulated clusters.

As a general result, both simulations and observational data show a
relative excess that increases with the projected radius. The lower
soft excess at small radii is due to the less complex thermal
structure of the ICM in the inner cluster regions: small sub--clumps
reaching such regions had time to be disrupted and their gas to get
thermalized with the surrounding hot cluster atmosphere. Still, for
$0.4\mincir R/R_{\rm vir}\mincir 0.7$, the excess observed in Coma is
somewhat higher than the level of the 90 percentile in the
distribution of the simulated soft X--ray excess. This indicates that
an excess as large as that observed in the Coma cluster represents a
rather rare event in our set of simulated clusters. We refrain from
drawing any strong conclusion from this comparison, since most of the
simulated clusters in our set are substantially smaller than the Coma
cluster.

\begin{center}
\begin{figure}
\psfig{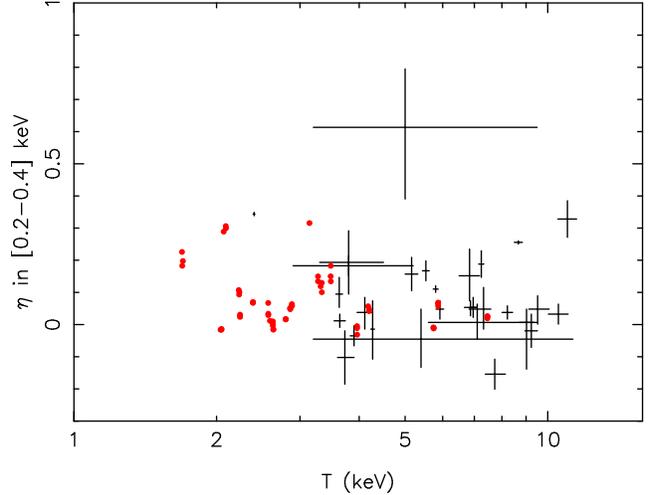} 
\caption{The relative excess $\eta$ in the [0.2-0.4] keV energy band
  for the observational data by Bonamente et al. (2002; crosses
  indicate the 1$\sigma$ uncertainties) and for simulations (filled
  circles). Results from simulations are obtained by computing the
  mean excess within $0.7\,R_{\rm vir}$, which represents a good
  approximation for the scales sampled by observations. As in Figure
  \ref{obs_coma}, we plot for each cluster the results obtained by
  projecting along three orthogonal directions.}
\label{observ}
\end{figure}
\end{center}

From their analysis of an ensemble of galaxy clusters, observed with
the ROSAT--PSPC, Bonamente et al. (2002) report that the soft excess
in the narrower [0.2--0.4] keV band is in fact a rather widespread
phenomenon, which is detected at high significance in a fair
fraction, $\sim 20$ per cent, of the sources. We compare in Figure
\ref{observ} the relative excess in this band as a function of the
cluster temperature, for both observed and simulated systems. As for
data, the values of the excess are those reported in Figure 3 by
Bonamente et al. (2002), while the temperature values have been taken
from White (2000). The simulation results are obtained by computing
the mean relative excess within $0.7\,R_{\rm vir}$, therefore
including also the contribution from the innermost regions. This scale
represents a good approximation to the typical size of the regions
covered by observations. Within the temperature interval sampled by
both observed and simulated clusters, the corresponding values of the
relative excess are quite consistent with each other. We note that few
observed systems have a significant negative excess. Bonamente et
al. (2002) interpret this as due to a soft absorption associated to
the presence of cold gas in central cluster regions. For $T\mincir 2$
keV the scatter in the relative excess increases for simulations, with
few clusters being characterized by a soft X--ray excess larger than
20 per cent. This indicates that the relative importance of the warm
component of the ICM increases in a few cases for colder systems, thus
potentially leading to an intrinsic difficulty of defining a single
temperature for systems with temperature below 2 keV (see also
Mazzotta et al. 2004).

As a concluding remark, it is worth mentioning that the approximate
treatment of metal production in our simulation represents a potential
source of uncertainty in the estimate of the soft excess. As already
mentioned, the simulation includes energy feedback and the production
of metals only from SN-II, using global metal yields (SH03). This is
the main reason for the rather low metallicity measured in the
simulated clusters (see Table 1). An improved treatment of the
chemical enrichment of the gas in simulations requires accounting for
the contribution from SN--Ia, hereby including detailed stellar yields
and stellar evolution models (e.g., Tornatore et al. 2004, and
references therein). As we have shown, the soft excess is associated
to clumped gas at a temperature of a few tenths of keV. At such
temperatures, a significant fraction of the emissivity is associated
to metal lines. Therefore, an underestimate of the plasma metallicity
would lead to an underestimate of the synthetic soft flux from the
warm clumps, while leaving nearly unaffected the soft flux from the
hot cluster atmosphere. In this respect, the values of the soft excess
presented here may be somewhat underestimated.

\section{Conclusions and future perspectives}

We have studied a sample of 20 simulated clusters, extracted from a
large SPH cosmological simulation of a concordance $\Lambda$CDM model,
with the aim of exploring the presence and the origin of the soft
X--ray excess in galaxy clusters. Besides the {\em non--radiative} gas
dynamics, our simulation includes the treatment of star formation and
energy feedback from galactic winds powered by supernovae. As such it
provides a quite realistic description of the evolution of the diffuse
gas in a cosmological environment.

Each cluster is observed in projection, by including the line-of-sight
contribution from the surrounding large-scale gas distribution. This
allows us to verify whether simulated clusters predict any soft excess
of thermal origin and whether such an excess is generated within or
outside the cluster virial region. Our analysis is aimed at mimicking
the observational procedure followed by Bonamente et al. (2002, 2003),
in their analysis of ROSAT--PSPC data. For each cluster we compute the
corresponding emission--weighted temperature and metallicity in a
relatively hard energy band. These values are then used to calculate
the spectrum of a one-temperature and one-metallicity plasma
model. The flux predicted by this spectrum in a softer band is then
compared to that from the synthetic spectrum, computed by summing over
the contribution of all the gas particles lying within the
``observational cylinder'' of each simulated cluster.

Our main results can be summarized as follows.
\begin{description}
\item[(a)] A significant soft X--ray excess is detected for
  cluster-centric distances $0.4<R/R_{\rm vir}<0.7$, whose amount
  varies from cluster to cluster (see Table 1 and
  Fig.\ref{fi:sp07}). In about 30 per cent of the cases we detect a
  relative excess at least as large as 50 per cent. However, the
  excess turns out to be always negligible in the central cluster
  regions, $R<0.4R_{\rm vir}$ (see Fig.\ref{fi:sp04}).
\item[(b)] In most of the cases, the excess is found to originate
  inside the virial region of the simulated clusters (see
  Fig.\ref{fi:exdis}). In the few cases in which a sizable
  contribution to the excess arises from the large--scale cluster
  environment, we find that it is contributed by fore/background
  groups lying along the cluster line-of-sight. Based on this result,
  we predict that the soft excess detected in observations receives
  only a minor contribution from the warm--hot medium permeating the
  large--scale cosmic web.
\item[(c)] Even within the virial cluster regions, the soft excess is
  contributed by high--density and low--entropy gas, rather than by a
  diffuse phase of warm gas (see Fig.\ref{fi:phase}). This gas is
  associated to merging sub--groups, which still preserve their
  identity before being destroyed and thermalized in the hot ICM. 
\item[(d)] We compare our results with those reported by Bonamente et
  al. (2003) for the Coma cluster. Although both observations and
  simulations show that the relative excess increases with the
  cluster-centric distance, its value in Coma is larger than for most
  of the simulated clusters. This implies that an excess as large as
  that observed for Coma is a rather rare event in our set of
  simulated systems.
\end{description}

A general conclusion of our analysis is that a soft X--ray excess of
thermal origin is naturally predicted by hydrodynamical simulations of
galaxy clusters in a cosmological environment. While this is an
interesting result {\em per se}, we believe that the comparison
between data and simulation does not yet have the required level of
accuracy to exclude the presence of a non--thermal origin (i.e.,
inverse Compton scattering), at least for part of the observed
excess. There is little doubt that solving this issue requires both a
more refined analysis of simulations and higher quality data.

Recent XMM--Newton observations have confirmed the presence of a soft
excess in the Coma cluster (Finoguenov et al. 2003) and in other four
clusters (Kaastra et al. 2003), which are consistent with having a
thermal origin. Finoguenov et al. (2003) claims that the soft excess
arises from a fairly large filamentary structure containing warm
gas. Although this result may be in contradiction with our results
from the simulations, it is clear that a detailed comparison would
require mimicking the observational procedure as close as possible on
a cluster-by-cluster basis.

It is clear that establishing the exact amount, nature and origin of
the soft X--ray excess requires an optimal control of observational
systematics, such as the instrumental calibration, accounting for
instrumental and cosmic background, as well as for the galactic
hydrogen column density in the cluster direction. Indeed, since the
soft emission occurs close to the lower boundary of the useful energy
range of currently available detectors, it is of paramount importance
to assess the accuracy of their calibration.

On the other hand, a detailed analysis of simulations requires
reproducing as close as possible the observational setup. In the
analysis presented in this paper, synthetic spectra of the simulated
clusters have been generated under the assumptions of dealing with an
ideal detector, with a flat response function across the whole energy
range of interest, and with no, or perfectly controlled,
background. It is clear that, with the steadily increasing quality of
data and level of sophistication of numerical simulations, accounting
for the instrumental features is mandatory for a self--consistent and
meaningful comparison between theoretical predictions and
observations, with both operating (e.g., Gardini et al. 2003) and
planned satellites (e.g., Yoshikawa et al. 2004). We are currently
undertaking a program of simulation analysis, based on mimicking
XMM--Newton spectrum observations with both pn and MOS detectors. This
will allow us to perform a more quantitative comparison with
observations, so as to shed light on the implication of the soft
excess on the physical properties of baryons within and around galaxy
clusters.

\begin{acknowledgements}
The simulation has been realized using the IBM-SP4 machine at the
``Consorzio Interuniversitario del Nord-Est per il Calcolo
Elettronico'' (CINECA, Bologna), with CPU time assigned thanks to an
INAF-CINECA grant. 
We are greatly indebted to Volker Springel, who provided the GADGET
code for the simulation and for a careful reading of the paper. 
We acknowledge useful discussions with Andrea
Biviano, Pasquale Mazzotta, Silvano Molendi and Xiang-Ping
Wu. L.-M.C. has been supported by a pre--doctoral fellowship of
Regione Friuli Venezia--Giulia. This work has been partially supported
by the PD51 INFN grant.

\end{acknowledgements}

\end{document}